\documentclass[10pt,letterpaper,twocolumn]{article} 

\usepackage{ol2}
\usepackage[draft]{hyperref}
\usepackage{amsmath}
\usepackage{setspace}
\begin{document}

\twocolumn[ 

\title{Fiber Based Multiple-Access Optical Frequency Dissemination}


\author{Y. Bai,$^{1,3}$ B. Wang,$^{1,2,*}$, X. Zhu,$^{1,3}$, C. Gao,$^{1,3}$ J. Miao,$^{1,3}$  and L. J. Wang$^{1,2,3,4*}$}

\address{
$^1$Joint Institute for Measurement Science, Department of Precision Instruments, Tsinghua University, Beijing 100084, China
\\
$^2$State Key Laboratory of Precision Measurement Technology and Instruments, Tsinghua University, Beijing 100084, China\\
$^3$Department of Physics, Tsinghua University, Beijing 100084, China\\$^4$National Institute of Metrology, Beijing 100013, China\\
$^*$Corresponding author: bo.wang@tsinghua.edu.cn, lwan@tsinghua.edu.cn
}

\begin{abstract}We demonstrate a fiber based multiple-access optical frequency dissemination scheme. Without using any additional laser sources, we reproduce the stable disseminated frequency at an arbitrary point along the fiber link. Relative frequency stabilities of $3\times10^{-16}/s$ and $4\times 10^{-18}/10^4s$ are obtained. A branching fiber network for high-precision synchronization of optical frequency is made possible by this method and its applications are discussed. \end{abstract}

 ] 

In recent years, to keep up with the significant progress of modern atomic frequency standard\cite{chou}, frequency dissemination via fiber link has been actively studied\cite{Warrington}. Different schemes, such as fiber based radio frequency (RF) dissemination\cite{Kumagai,Wang}, optical frequency dissemination\cite{Williams,Lopez,Predehl}, and even optical frequency comb signal dissemination\cite{Marra,Hou} have been proposed and demonstrated, respectively. All of these schemes have much higher dissemination stability than that of the conventional frequency dissemination via satellite link, which has a limited stability of $10^{-15}$/day\cite{Bauch,Levine}. Ultra-long distance frequency dissemination has also been demonstrated by cascaded frequency dissemination method\cite{Predehl,Fujieda2,Lopez2}. However, currently, almost all of the frequency dissemination fiber links are built between precision laboratories and used for atomic clock comparisons\cite{Ludlow, Hong}. Most of the application areas, such as distributed synthetic aperture radar\cite{radar}, Very-long-baseline interferometry (VLBI)\cite{VLBI}, and particle accelerators\cite{Norman}, still use satellite links to synchronize the time and frequency. It is mainly due to the frequency accessing limitation of the fiber link. For satellite link, it can disseminate time and frequency signal to cover essentially the entire globe; for fiber link, the point to point frequency dissemination protocol limits its coverage area.

Recently, we have proposed and demonstrated a multiple-access RF dissemination method\cite{Gao}. Using it, one can access the high-quality frequency signal at any point along an existing RF dissemination fiber link, which makes it possible to build a branching RF synchronization fiber network. In 2010, for the first time, G. Grosche of PTB proposed a scheme for multiple-access optical frequency dissemination\cite{Grosche}. In this scheme, a laser system with almost same frequency as that of the disseminated light is required at the accessing point. Furthermore, the long term frequency drift of the laser system can not exceed the working bandwidth of the compensating AOM (normally several tens of MHz). These two requirements make the proposed scheme difficult to realize at an arbitrary remote point along the fiber link. Maybe it is the reason why there are no experimental demonstrations for this scheme till now. In this letter, partly based on Grosche's scheme, we demonstrate a multiple-access optical frequency dissemination scheme without using any additional laser source. As a preliminary demonstration, we disseminate a 1550 nm laser light via a 3 km fiber link which is composed by two 1.5 km fiber spools.  The arbitrary frequency accessing point is chosen as being 1.5 km away from the sending site. Relative frequency stabilities of $3\times10^{-16}/s$ and $4\times 10^{-18}/10^4s$ for the distributed optical frequency are obtained, which are at the same level as the dissemination stabilities of the entire fiber link ($1\times10^{-16}/s$ and $1\times 10^{-18}/10^4s$). Here, the technique can be equally applied to longer fiber link given a laser with a narrower linewidth. With this multiple-access capability, the fiber-based optical frequency dissemination system will become more applicable as a time and frequency network. The existing ultra-long distance optical frequency dissemination project\cite{Predehl} and many other application areas\cite{radar,VLBI, Norman} may benefit from this work.

\begin{figure}[htb]
\centerline{\includegraphics[width=8.5 cm]{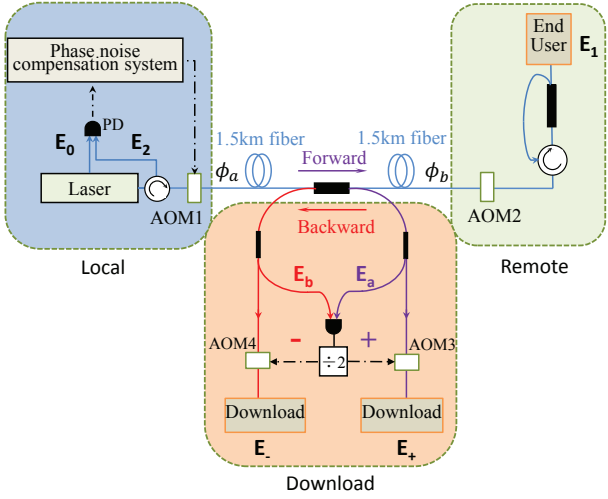}}
\caption{(Color online) Schematic diagram of the fiber based multiple-access optical frequency dissemination system.}
\label{fig1}
\end{figure}

 Figure~\ref{fig1} shows the schematic diagram of the fiber based multiple-access optical frequency dissemination experiment. The whole system is separated into three parts, which are the local site, remote site and download site, respectively. For the convenience of phase stability measurement, they are located in the same lab. As a performance test, a commercial 1550nm laser module (RIO ORION Laser Module) is used as the reference resource at the local site. Its  frequency signal can be expressed as $E_0=V_0cos(\omega t+\phi_0)$. After passing a circulator and an 80MHz fiber-couple acoustic optical modulator AOM1 (+1st order Bragg shift), the laser signal is coupled into the fiber link. A phase fluctuation $\phi_p$ of the signal $E_0$ will be induced during its propagation. At the remote site, another 80MHz AOM2 (+1st order Bragg shift) is employed as a frequency shifter for optical beating at the download site. The received signal $E_1$ at the remote site can be expressed as \begin{equation}
E_1=V_1cos(\omega t+\Omega_1t+\Omega_2t+\phi_0+\phi_p+\phi_{AOM1}).
\end{equation} Here, $\Omega_1\approx\Omega_2\approx2\pi\times$80 MHz are the frequency offsets induced by AOM1 and AOM2, $\phi_{AOM1}$ is the correction phase modulated on AOM1, and the fixed phase shift induced by AOM2 is neglected. To compensate the phase fluctuations induced during fiber dissemination, at remote site, a fraction of the received optical signal is sent back (via an optical circulator) to beat with $E_0$ at the local site. The round-trip signal $E_2$ can be expressed as
 \begin{equation}
E_2=V_2cos(\omega t+2\Omega_1t+2\Omega_2t+\phi_0+2\phi_p+2\phi_{AOM1}),
\end{equation} and the beating signal is proportional to $cos(2\Omega_1t+2\Omega_2t+2\phi_p+2\phi_{AOM1})$. The beating signal is processed by the phase noise compensation system, and the generated error signal is fed to control the phase of AOM1. It is the same way as that of conventional optical-stabilization techniques\cite{Ma}. Once the phase-locking loop is closed, without loss of generality, we can get:
\begin{equation}
\phi_p+\phi_{AOM1}=0.
\label{eq1}
\end{equation}
Consequently, the phase noise compensation is accomplished and the received frequency signal $E_1$ is phase locked to $E_0$ at the local site.

\begin{figure}[htb]
\centerline{\includegraphics[width=8cm]{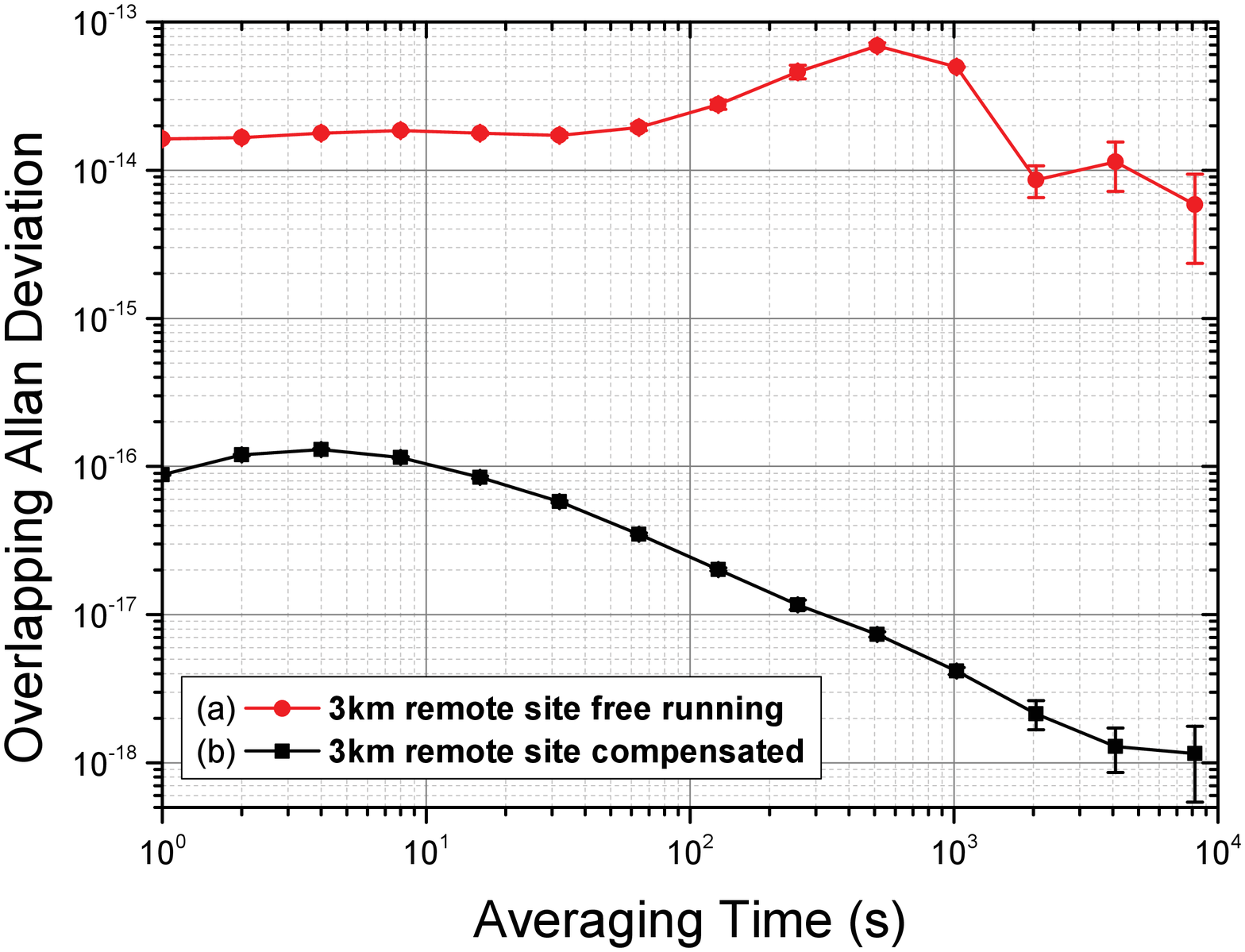}}
\caption{(Color online) Measured relative frequency stability of the received optical frequency $E_1$ at the remote site when (a) the 3km fiber link running freely, and (b) the phase noise is actively compensated.}
\label{fig2}
\end{figure}

Through beating $E_0$ and $E_1$, we measure the dissemination stability of the compensated fiber link. Figure~\ref{fig2} shows the measurement results. Using the phase noise compensation system, dissemination stabilities of $1\times10^{-16}/$s and $1\times10^{-18}/10^4s$ are achieved. While for the free-running fiber link, it have dissemination stabilities of $1\times10^{-14}/$s and $6\times10^{-15}/10^4s$, respectively.

\begin{figure}[htb]
\centerline{\includegraphics[width=8cm]{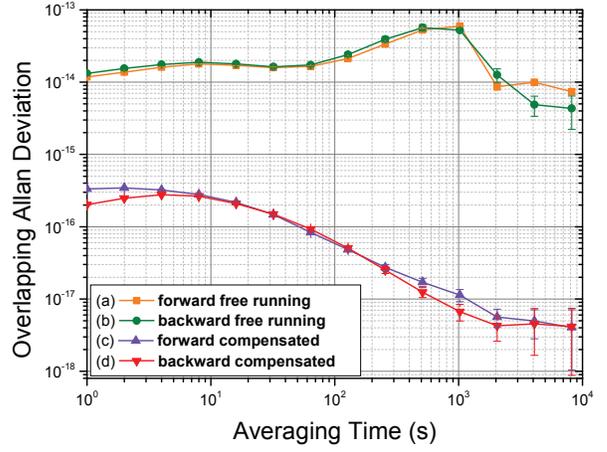}}
\caption{(Color online) Measured relative frequency stabilities of frequency signals at the download site with and without phase noise compensated. (a) forward signal directly detected; (b) backward signal directly detected; and (c) forward signal with compensation; (d) backward signal with compensation.}
\label{fig4}
\end{figure}

In order to download the disseminated frequency signal at an arbitrary accessing point, using a $2\times2$ fiber coupler, we can easily couple out the forward and backward propagating optical signal from the fiber link. Due to the existence of AOM2, there is a frequency difference of 160 MHz between the two signals. Similarly, they can be expressed as
\begin{subequations}
\begin{align}
E_a&=V_acos(\omega t+\Omega_1t+\phi_0+\phi_a+\phi_{AOM1}),\\
E_b&=V_bcos(\omega t+\Omega_1t+2\Omega_2t+\phi_0+\phi_p+\phi_b+\phi_{AOM1}).
\end{align}
\end{subequations}
Here, $\phi_a$ and $\phi_b$ are the phase fluctuations induced by the fiber links in front of and behind the accessing point, respectively. They obviously have the relationship of $\phi_a+\phi_b=\phi_p$.  By beating  $V_a$ and $V_b$ with $V_0$ respectively, we can get the relative frequency stabilities of the directly accessing signals. As shown in Fig.~\ref{fig4}(curve (a), (b)), they are $1\times10^{-14}/s$, $6\times 10^{-15}/10^4s$ for the forward signal and   $1\times10^{-14}/s$, $4\times 10^{-15}/10^4s$ for the backward signal. The noise levels are similar to that of a free running fiber link (curve (a) of Fig.~\ref{fig2}). To cancel the phase fluctuations $\phi_a$ and $\phi_b$ of the two directly accessed optical signals, we simply beat a small fraction of the forward and backward transmitted signals $E_a$ and $E_b$, and get
\begin{equation}
V_{beat} \propto cos(2\Omega_2t+\phi_p+\phi_b-\phi_a)=cos(2\Omega_2t+2\phi_b).
\end{equation}

Two more accousic optical modulators AOM3(+1st order Bragg shift) and AOM4(-1st order Bragg shift) are further employed as shown in Fig.~\ref{fig1}. The beat signal $V_{beat}$ is discriminated and digitally divided by 2. The resulting 80MHz RF signal with phase $\phi_b$ is then bandpass filtered, amplified, and fed to AOM3 and AOM4. The modulated, download signals of the forward and backward directions can be written as:

\begin{subequations}
\begin{align}
E_+&=V_+cos(\omega t+\Omega_1t+\Omega_2t+\phi_0+\phi_a+\phi_b+\phi_{AOM1}),\\
E_-&=V_-cos(\omega t+\Omega_1t+\Omega_2t+\phi_0+\phi_p+\phi_{AOM1}).
\end{align}
\end{subequations}
When the phase noise compensation loop is closed as described above,  considering Eq.~\eqref{eq1}, the signals reproduced at the download site become:
\begin{equation}
E_+=E_-\propto cos(\omega t+\Omega_1t+\Omega_2t+\phi_0).
\end{equation}

Therefore, the reproduced signals $E_+$ and $E_-$ have exactly the same form as $E_1$ received at the remote site. As shown in Fig.~\ref{fig4}(curve (c), (d)),  relative frequency stabilities can be improved to $3\times10^{-16}/s$, $4\times 10^{-18}/10^4s$ for the forward signal, and $2\times10^{-16}/s$, $4\times 10^{-18}/10^4s$ for the backward signal. We note that the relative frequency stabilities of download signals are slightly worse compared with the dissemination stability of the whole fiber link. This is mainly caused by the phase fluctuations induced by the out-of-loop fibers and AOMs at the download site. By thermally and acoustically isolating the out-of-loop elements, more stable optical frequency signal can be downloaded. However, as a preliminary demonstration, we emphasize that the advantages of this method, namely, there is no requirement of any additional laser source at the download site, and the apparatus is as simple as that of remote site. This feature will greatly improve the feasibility of a branching fiber network for optical frequency dissemination.

In summary, we have demonstrated a fiber-based, multiple-access optical frequency dissemination scheme. Using the method, stable, distributed optical frequency signals can be accessed conveniently at an arbitrary point along the entire fiber link. The relative frequency stability is at the same level with the dissemination stability of the whole fiber link. A branching fiber network for high-precision synchronization of optical frequency is thus enabled by this method, possibly expanding applications of the existing ultra-long distance optical frequency dissemination project into other application areas.

The authors acknowledge funding supports from the Major State Basic Research Development Program of China (No.2010CB922901) and Tsinghua University Scientific Research Initiative Program (No.20131080063).

\pagebreak
\section*{Informational Fourth Page}
 {\bf Full versions of citations}

\end{document}